\documentclass[12pt,preprint]{aastex}

\def\gta{\ifmmode {\mathbin{\lower 3pt\hbox   
    {$\,\rlap{\raise 5pt\hbox{$\char'076$}}\mathchar"7218\,$}}}
    \else {${\mathbin{\lower 3pt\hbox
    {$\rlap{\raise 5pt\hbox{$\char'076$}}\mathchar"7218\,$}}}
    $}\fi}
\def\lta{\ifmmode {\,\mathbin{\lower 3pt\hbox   
    {$\,\rlap{\raise 5pt\hbox{$\char'074$}}\mathchar"7218\,$}}}
    \else {${\mathbin{\lower 3pt\hbox
    {$\rlap{\raise 5pt\hbox{$\char'074$}}\mathchar"7218\,$}}}
    $}\fi}

\begin{document}

\title{Gravitational Radiation from Intermediate-Mass Black Holes}

\author{M. Coleman Miller}
\affil{Department of Astronomy, University of Maryland\\
       College Park, MD  20742-2421\\
       miller@astro.umd.edu}

\begin{abstract}

Recent X-ray observations of galaxies with ROSAT, ASCA, and Chandra have
revealed numerous bright off-center point sources which, if isotropic
emitters, are likely to be intermediate-mass black holes, with $M\sim
10^{2-4}\,M_\odot$.  The origin of these objects is under debate, but
observations suggest that a significant number of them currently reside
in young high-density stellar clusters.  There is also growing
evidence that some Galactic globular clusters harbor black holes of similar 
mass, from observations of stellar kinematics.
In such high-density stellar environments, the interactions of
intermediate-mass black holes are promising sources of gravitational
waves for ground-based and space-based detectors. Here we explore
the signal strengths of binaries containing intermediate-mass black
holes or stellar-mass black holes in dense stellar clusters.  
We estimate that a few to tens per year of these objects will be
detectable during the last phase of their inspiral with the advanced LIGO
detector, and up to tens per year will be seen during merger,
depending on the spins of the black holes.  We also find that if these
objects reside in globular clusters then tens of sources will be
detectable with LISA from the Galactic globular system in a five year
integration, and similar numbers will be detectable from more distant
galaxies.  The signal strength depends on the eccentricity distribution,
but we show that there is promise for strong detection of pericenter
precession and Lense-Thirring precession of the orbital plane.  We
conclude by discussing what could be learned about binaries, dense
stellar systems, and strong gravity if such signals are detected.

\end{abstract}

\keywords{black hole physics --- gravitational waves --- stellar dynamics}

\section{Introduction}

We are entering an era in which numerous experiments will search for
astrophysical gravitational radiation.  Ground-based detectors such as
LIGO (Barish 2000), VIRGO (Fidecaro et al. 1997), GEO600  (Schilling
1998), TAMA (Ando et al. 2002), and others focus on relatively high
frequencies $f_{\rm GW}\approx 10-1000$~Hz.  These frequencies are
appropriate for the final inspiral and merger of binaries with total
masses $\sim 1-1000\,M_\odot$. Space-based missions such as the Laser
Interferometer Space Antenna (LISA; see, e.g., Danzmann 2000) will
complement this frequency range by focusing on comparatively low
frequencies, $f_{\rm GW}=10^{-4}-1$~Hz. In this range there are a number
of ``guaranteed" sources in the form of main sequence and white dwarf
binaries, but there is much  more uncertainty about the prevalence and
properties of double compact binaries containing neutron stars and black
holes.  Such binaries are of great interest for many reasons, such as
their use as probes of strong gravity and their potential to illuminate
other aspects of astronomy, including the evolution of galaxies.

The existence of a new class of double compact binaries is suggested by
two  recent lines of evidence.  First, observations of the kinematics of
the central regions of some globular clusters suggests that black holes
with $M\gta 100\,M_\odot$ may exist in their cores (e.g., Gebhardt et al.
2000; D'Amico et al. 2002).  Second, X-ray observations have revealed
variable, unusually high flux point sources in a number of galaxies
(e.g., Fabbiano 1989; Fabbiano, Schweizer, \& Mackie 1997; Colbert \&
Mushotzky 1999; Zezas, Georgantopoulos, \& Ward 1999; Kaaret et al. 2001;
Matsumoto et al. 2001; Fabbiano, Zezas, \& Murray 2001; see Colbert \&
Ptak 2002 for a catalog of objects).  The rapid and strong variability of
these sources indicates that they are black holes.  If their flux is not
strongly beamed, then for their luminosities to be below the Eddington
luminosity $L_E=1.3\times 10^{38}(M/M_\odot)$~erg~s$^{-1}$, at which
radiation forces balance gravity, the masses of the brightest sources
must be at least $\sim 10^3\,M_\odot$ (e.g., Matsumoto et al. 2001).  In
addition, their off-center positions in their host galaxies indicate that
in many cases the masses cannot be substantially greater than $\sim
10^5\,M_\odot$, otherwise dynamical friction would cause the black holes
to sink rapidly to the dynamical center (Kaaret et al. 2001; Tremaine,
Ostriker, \& Spitzer 1975). Recent optical observations suggest that for
at least some of the sources, the X-rays may not be beamed strongly,
based on recombination emission from surrounding nebulae (Pakull \&
Merioni 2002). This implies that the total luminosity is indeed high, so
that if the sources are sub-Eddington (but see Begelman 2002) there may
therefore be a significant number of $\sim 10^{2-4}\,M_\odot$ black holes
in the universe. Currently, these sources are preferentially in star
forming regions or in starburst galaxies (Matsumoto et al. 2001; Liu \&
Bregman 2001).

A number of these objects are associated with dense young stellar clusters
(e.g., Matsumoto et al. 2001).  Combined with the kinematic evidence in
globular clusters, this implies that black holes of this sort undergo
frequent dynamical encounters.  It has been suggested that the ultraluminous
X-ray sources originated due to dynamical encounters in globular clusters,
then merged with their host galaxies (Miller \& Hamilton 2002a,b).  Other
proposals, if these objects are $10^{2-4}\,M_\odot$ black holes instead of
being strongly beamed (King et al. 2001; King 2002; K\"ording, Falcke,
\& Markoff 2002) or super-Eddington by a
large factor (Begelman 2002), are that they form as a result of a core
collapse of a massive young star cluster (Matsushita et al. 2000; Ebisuzaki
et al. 2001) or as remnants of Population~III stars (Madau \& Rees 2001).

Here we consider the implications for gravitational wave emission if these
sources are indeed intermediate-mass black holes.  Independent of  their
detailed origin, black holes in dense stellar environments are promising
sources of gravitational radiation due to the diversity of dynamical
interactions that are possible (see also Benacquista 1999, 2002a,b;
Benacquista, Portegies Zwart, \& Rasio 2001).   We will focus mainly on
black holes in globular clusters, but will also address signals that may be
evident from young stellar clusters, regardless of whether the black hole
formed there or elsewhere.  In \S~2 we give general arguments about the
properties of these sources, based on the ages of globulars and the number
and types of secondaries.  In \S~3 we estimate the rates of encounters, and
in \S~4 we present calculations of the signal to noise ratios expected for
LISA from Galactic globulars and from globulars in the Virgo cluster, and
for coalescence of black holes as observed with the LIGO-II detector.  In
\S~5 we discuss what information could be gleaned from a reasonably high
signal to noise detection.  We show that both pericenter precession and
Lense-Thirring precession are potentially measurable.  These effects may
allow preliminary mapping of the spacetime around black holes and could
even yield an independent measure of the distance to the Virgo cluster.
We present our conclusions and discuss the uncertainties in our
estimates in \S~6.

\section{Expected Properties of Sources}

Here we discuss the expected nature of intermediate-mass black holes
if these are grown in dense clusters.  We begin by summarizing the
dynamics of dense stellar clusters and then discuss the ways in which
intermediate-mass black holes can form and grow.  We then derive the
approximate distributions of spin parameter and eccentricity that we
expect for these sources, with implications for detectability.

\subsection{Cluster dynamics}

The young super star clusters now being discovered in many galaxies with
active star formation (e.g., Origlia et al. 2001) have typical estimated
masses  $\sim 10^6\,M_\odot$ and half-mass radii 10~pc (Origlia et al.
2001).  The central number density is difficult to estimate, but if we
take as a guide the densest young clusters in the vicinity of the Milky
Way the central density could be up to $\sim 10^{5-6}$~pc$^{-3}$ (see,
e.g.,  Massey \& Hunter 1998 for observations of the dense R136
region).   Clusters with ages less than a few tens of millions of years
have a large number of O and B stars; older than this, and those stars
have evolved to compact remnants.

Globular clusters are much older, $\sim 10^{10}$~yr.  All stars with
initial masses $M_{\rm init}\gta 0.8\,M_\odot$ have evolved off the main
sequence, meaning that the most massive objects present are compact
remnants such as black holes, neutron stars, and massive white dwarfs.
There are also a small number of $\sim 1.5\,M_\odot$ blue stragglers,
which are main sequence stars more massive than the ordinary main-sequence
turnoff that may have been rejuvenated by collisions (Lombardi et al.
2002; see, e.g., Hurley \& Shara 2002 for a recent discussion of these and
other interactions in clusters). A typical globular has $\sim 10^{5-6}$
stars of average mass  $\sim 0.4\,M_\odot$, and a central number density
in the range $n_c=10^2-10^6$~pc$^{-3}$ (Pryor \& Meylan 1993).  Roughly
40\% of the globulars surrounding the Milky Way have $n_c\gta
10^5$~pc$^{-3}$ (Pryor \& Meylan 1993), with a density contrast between
the core and the half-mass radius that formally poise them on the edge of
core collapse. It is thought that three-body interactions of primordial
binaries with field stars heat the cluster, delaying collapse much
longer than otherwise possible (e.g., Goodman \& Hut 1989;
Hut, McMillan, \& Romani 1992; Sigurdsson \& Phinney 1995).

A productive analogy with thermodynamics, backed up by simulations, has
shown that in clusters with stars of different masses the cluster tries
to evolve towards thermodynamic equilibrium, in the sense that the
kinetic energies of each of the component stars are drawn from the same
distribution (see, e.g., Binney \& Tremaine 1987).  This means that the
typical speed of an object of mass $M$ is proportional to $M^{-1/2}$.
This causes more massive objects to sink to the core of the cluster, a
process called mass segregation (e.g., Farouki \& Salpeter 1982).
The typical timescale for such sinking is given by
$t\approx t_r(\langle m\rangle/M)$, where for a cluster of $N$ stars and
dynamical crossing time $t_d$ the core relaxation timescale is
$t_r\approx t_dN/(8\ln N)$ (see, e.g., Binney \& Tremaine 1987).
Typical globulars have $t_r\approx 10^{7-9}$~yr (Pryor \& Meylan 1993).
This means that even small black holes with $M=10\,M_\odot$ will sink
quickly to the center of a cluster, and if an intermediate-mass black
hole with $M=10^{2-4}\,M_\odot$ is present it will rarely be too far
from the center even in a young cluster.  However, the short relaxation
time in the core and wandering of black holes of this mass due to
dynamical interactions mean that there is rapid refilling of stellar
orbits that interact with the black hole.  There is therefore no ``loss
cone" problem, unlike the situation with supermassive black holes in
galactic nuclei (Frank \& Rees 1976; see Milosavljevi\'c \& Merritt 2001
for a recent discussion).

As a result of mass segregation, even though the absolute number of
compact objects is much less than that of main sequence stars, their
number density in the core can be comparable to or in excess of that of
main sequence stars for core densities $\gta 10^5$~pc$^{-3}$ (e.g.,
Sigurdsson \& Phinney 1995).  In addition, since binaries typically
have more mass than single stars, they also tend to sink to the center.
This means that three-body interactions of binaries with single stars,
and possibly binary-binary interactions, can be dynamically important.

Qualitatively, three-body interactions can be understood using Heggie's
rule (Heggie 1975) that hard binaries harden and soft binaries soften.
That is, if the initial binary has a binding energy greater than the
typical kinetic energy of a field star (i.e., the binary is hard) then
a typical encounter with a field star will cause the binary to harden
further.  A soft binary will instead be softened gradually until it
becomes unbound.  Exchanges can also occur, in which the interloper
object becomes a member of the final binary.  Numerous simulations show
that in a strong encounter the most likely occurrence is that the final
binary is composed of the two most massive of the three stars that
originally interacted (e.g., Sigurdsson \& Phinney 1993; Heggie, Hut,
\& McMillan 1996).

There has long been interest in whether hardening from three-body encounters
might bring a pair of black holes close enough together that they would
merge via gravitational radiation while still in the cluster. However, each
hardening delivers a recoil kick to both the binary and the
single star, and if the recoil of the binary exceeds the $\sim
50$~km~s$^{-1}$ escape speed from the core of a typical dense cluster
(Webbink 1985) then the binary will be ejected before it can merge.  Several
simulations (Kulkarni, Hut, \& McMillan 1993; Sigurdsson \& Hernquist 1993;
Portegies Zwart \& McMillan 2000) have shown that if attention is restricted
to Newtonian three-body encounters of $10\,M_\odot$ black holes then few if any
mergers will occur inside the cluster. There may, however, be a number of
mergers that happen outside the cluster after recoil (Portegies Zwart \&
McMillan 2000).

Recently, Miller \& Hamilton (2002a) showed that merger may happen inside
the cluster if a black hole starts out with a greater mass, $M\sim
30-50\,M_\odot$.  In this case, recoil is much less significant and
gravitational radiation predominates.  They proposed this as a mechanism
for the formation and growth of intermediate-mass black holes.  Even if
only lighter black holes form initially, secular resonances in
hierarchical triple systems formed in binary-binary encounters (the Kozai
resonance: Kozai 1962; Lidov \& Ziglin 1976) may increase the eccentricity
of the inner binary enough that it can merge by gravitational radiation
without being ejected by three-body recoil (Miller \& Hamilton 2002b).
It is also possible that in some three-body encounters, two of the black
holes will pass close enough to each other to coalesce rapidly.
This increases optimism that dense clusters may be the seat of relatively
frequent mergers of compact objects.  There are, however, a number of
questions still remaining.  One of these has to do with recoil due to
asymmetric emission of gravitational radiation during inspiral.
Calculations have yet to be done for strong gravity, but various analyses
using Newtonian and post-Newtonian formalisms suggest that the recoil
velocity could be from a few kilometers per second up to $\sim
10^3$~km~s$^{-1}$, depending on the mass ratio and how close the black
holes get before the final plunge (Peres 1962; Bekenstein 1973; Fitchett
1983; Fitchett \& Detweiler 1984; Redmount \& Rees 1989; Wiseman 1992).
For slowly-rotating black holes with mass ratios $M/m\gta 10$, the prime
focus of this paper, the best estimates are that the recoil speed will be
well below the 50~km~s$^{-1}$ escape speed from the core (Fitchett 1983;
Fitchett \& Detweiler 1984; Wiseman 1992). We will therefore assume that
if an intermediate-mass black hole forms in a dense cluster it will remain
in the core through multiple mergers.

\subsection{Spin parameter}

Black holes that have grown by the accumulation of smaller objects in a
dense stellar cluster are likely to have a different rotation parameter
than either stellar-mass or supermassive black holes. Stellar-mass black
holes acquire some spin at birth.  Subsequent accretion adds relatively
little mass or angular momentum to the black hole, either because not much
mass is available (as in low-mass X-ray binaries, where the companion is
usually less than 10\% as massive as the black hole; e.g., Verbunt 1993)
or because the duration of the accretion phase is short (as in high-mass
X-ray binaries, with durations $T<10^7$~yr implying a total mass transfer
less than $0.1\,M_\odot$: Verbunt 1993).  As a result, the spin parameter
$j\equiv cJ/GM^2$, where $c$ is the speed of light, $G$ is Newton's
gravitational constant, and $J$ is the angular momentum, of stellar-mass
black holes is expected to be close to its birth value. Compelling
evidence of significant spin in stellar-mass black holes has emerged from
study of quasi-periodic brightness  oscillations (Strohmayer 2001a,b;
Miller et al. 2001), and spectral studies are also suggestive of high spin
(from line profiles,  Miller et al. 2002; and earlier more model-dependent
fits  to continuum emission, Zhang, Cui, \& Chen 1997).  For supermassive
black holes, the spin parameter depends on what fraction of its mass was
provided by an accretion disk with a fixed orientation.  If the fraction
is close to unity then the spin parameter could also be close to unity.
Some evidence of such rapidly spinning supermassive black holes is
emerging in the form of extremely broad Fe~K$\alpha$ lines from a few
active galactic nuclei (e.g., Iwasawa et al. 1996, Dabrowski et al. 1997;
Wilms et al. 2001).

In contrast, black holes grown by the capture of stars or compact objects
in dense stellar clusters undergo a damped random walk in the evolution of
their spins after the memory of the initial spin is lost.  The damping is
because retrograde orbits become unstable at larger specific angular
momentum than do prograde orbits, so that it is easier to decrease than to
increase the spin parameter of the black hole.  Assuming a roughly
isotropic distribution of stellar velocities in the cluster core, the final
inspiral and deposition of angular momentum will happen at random angles to
the previous spin axis of the black hole.  There is
evidence of net rotation in many globular clusters (e.g., Barmby et al.
2002), so that encounters may not be precisely isotropic, but this is
likely to be a small effect because relaxation will drive the core
distribution towards isotropy (e.g., Einsel \& Spurzem 1999; Kim et al.
2002 for Fokker-Planck treatments of the evolution of rotating systems).

If the angular momentum of the large black hole is much greater
than the orbital angular momentum of a small black hole that falls
into it, then the total angular momentum is roughly constant, so
that $j\sim M^{-2}$ (Hughes \& Blandford 2002).  In contrast,
if the angular momentum of the larger black hole is sufficiently
small, then the orbital angular momentum of the smaller black hole
can make a significant difference.  Following Hughes \& Blandford (2002),
let us define the orbital inclination $\iota$ using
\begin{equation}
\cos\iota={L_z\over{\sqrt{L_z^2+Q}}}\; ,
\end{equation}
where $L_z$ is the angular momentum parallel to the spin axis of
the more massive black hole and $Q$ is the Carter constant.
If the initial spin angular momentum of the massive black hole is
$J$, then merger with a black hole of mass $m\ll M$
with an orbital inclination $\iota$ and specific orbital angular momentum 
$u_\phi(\iota)$ will produce a black hole of mass $M+m$ and
angular momentum
\begin{equation}
\begin{array}{rl}
J^\prime&=\sqrt{(J+mu_\phi(\mu)\cos\iota)^2+(1-\cos^2\iota)m^2
u_\phi^2(\iota)}\\
&=\sqrt{J^2+2mJu_\phi(\iota)\cos\iota+m^2u_\phi^2(\iota)}\;.\\
\end{array}
\end{equation}
This assumes that the amount of mass-energy and angular momentum 
radiated away during the merger is small.  Hughes \& Blandford (2002)
show that to good accuracy the angular momentum of a test particle
in the last stable circular orbit 
at angle $\iota$ to the prograde equatorial direction is
approximately
\begin{equation}
|u_{\rm \phi,LSO}(\iota)|\approx |L_{\rm ret}|+{1\over 2}(\cos\iota +1)
\left(L_{\rm pro}-|L_{\rm ret}|\right)\; ,
\end{equation}
where $L_{\rm pro}$ and $L_{\rm ret}$ are, respectively, the 
specific angular momentum of a prograde equatorial and retrograde
equatorial particle at the last stable orbit.  When $j\ll 1$ this
reduces to
\begin{equation}
|u_{\phi,LSO}|\approx\sqrt{12}M(G/c)\left[1-{1\over 2}
\left(2\over 3\right)^{2/3}j\cos\iota\right]\; ,
\end{equation}
Assume the probability distribution of inclinations is $Q(\cos\iota)$
(e.g., $Q=1/2$ for an isotropic distribution).  If the probability
distribution for $j$ given $M$, $P(j|M)$, is stationary in form, then
we have
\begin{equation}
P(j|M+m)=\int_{-1}^1P\left[j+\delta(\cos\iota)|M\right]
Q(\cos\iota)\,d(\cos\iota)\; ,
\end{equation}
where $\delta(\cos\iota)$ is the change in dimensionless angular
momentum caused by the accretion of an object of mass $m$ at an
orbital inclination $\iota$; that is, such an object will change
the dimensionless angular momentum from $j+\delta(\cos\iota)$ to $j$.

Approximate solutions to this equation may be sought in the form
$P(j|M)\approx \exp[-(j-\beta)^2/2\sigma^2]$.  For isotropic encounters,
a fit to the numerical results displayed in Figure~1 gives $\beta\approx
(2m/M)^{1/2}$ and $\sigma\approx (m/2M)^{1/2}$. Figure~1 shows the
distribution of spin parameters for a number of simulated black holes
that started at $M_0=5m$ and accreted objects of mass $m$ (e.g.,
$50\,M_\odot$ and $10\,M_\odot$).  For intermediate-mass black holes,
the expected dimensionless angular momentum is on the order of a few
tenths, approximately independent of the initial spin if $M_0\geq 5m$.

\centerline{\plotone{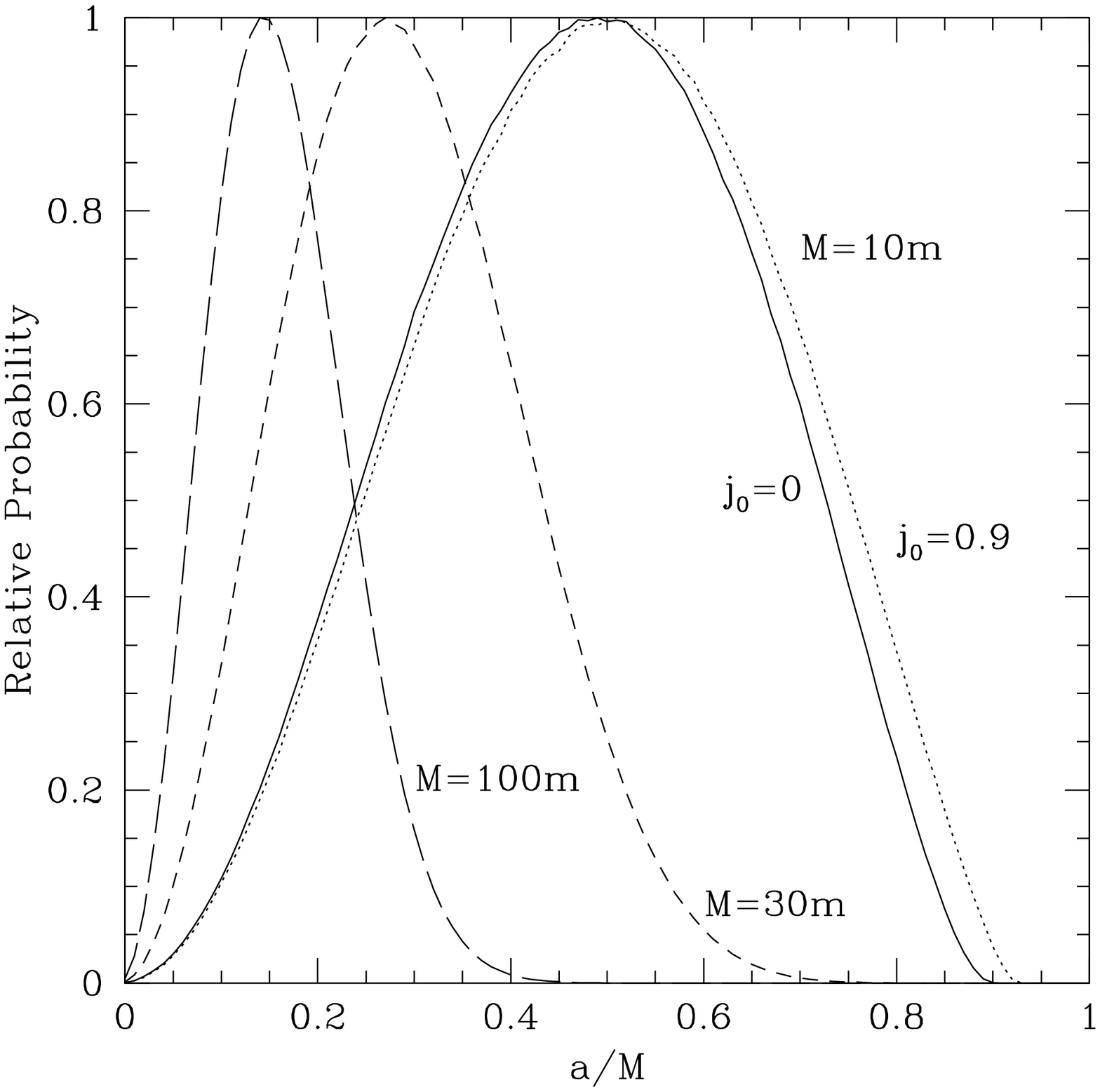}}
\figcaption[]{Simulated distribution of spin parameters $j=a/M=cJ/GM^2$ for
a black hole of initial mass $5m$ that accretes
objects of mass $m$ up to a total of $10m$ (solid line:
initial spin $j=0$; dotted line: initial spin $j=0.9$),
$30m$ (short dashed line; initial spin $j=0$), or
$100m$ (long dashed line; initial spin $j=0$).  Orbital
inclinations $\iota$ of encounters are uniformly distributed in
$\cos\iota$.}

\subsection{Eccentricity}

The expected eccentricity for a black hole binary depends on
the way in which it was formed and on the orbital frequency at
which it is observed.  We will discuss each of the evolutionary
paths in turn, but in brief we find that at the high frequencies
observable with ground-based detectors the orbits will have
circularized to high accuracy, whereas at the lower frequencies
of space-based detectors the eccentricity is expected to be significant
for the majority of the detectable period.

First, consider formation via three-body encounters.  As long as the time
to merger is longer than the time to the next encounter, the eccentricity
will be reset by each interaction.  During this period there will be
relatively little gravitational radiation emitted, as the merger time is
typically long.  At some point, however, an encounter leaves the binary
with a relatively short merger time and thus the binary goes into
uninterrupted two-body inspiral.  The precise distribution of
eccentricities at the start of this phase depends on the details of the
three-body interactions, but preliminary results (Gultekin \& Miller, in
preparation) suggest that immediately after the last three-body encounter the
eccentricity tends to be fairly high, $e>0.9$.  This is because the time
to merger decreases rapidly with increasing eccentricity, so a chance
fluctuation up to high eccentricity will allow a binary to merge before
the next encounter.  For moderate to high eccentricities the gravitational 
radiation merger time is 
\begin{equation}
\tau_{\rm merge}\approx 3\times 10^{17}(M_\odot^3/\mu M^2)
(a/1~{\rm AU})^4(1-e^2)^{7/2}\,{\rm yr}
\end{equation}
(Peters 1964), where $M$ is the total mass and $\mu$ is the reduced mass
of the binary.  For $e\approx 1$ this is $\tau_{\rm merge}\approx 3\times
10^{18}(a/1~{\rm AU})^{1/2}(r_p/1~{\rm AU})^{7/2}$~yr, where $r_p=a(1-e)$
is the pericenter distance.  For a typical merger time, after the last
encounter, of a few hundred thousand to a few million years, the
pericenter distance is therefore $r_p\sim {\rm few}\times 10^{10}$~cm,
which is hundreds to thousands of gravitational radii for an
intermediate-mass black hole.  Although the full period of an orbit could
be relatively high ($10^{4-5}$~s), the majority of the power is emitted at
a frequency of the order of the angular frequency at pericenter (see
\S~4.1), which could be several times higher than the fundamental
orbital frequency.  The peak frequency of gravitational
waves is then typically on the order of $10^{-4}-10^{-3}$~Hz for the
majority of the inspiral.  This is what would most likely be seen from
sources in our Galactic globular system, and hence the eccentricity would
be expected to be high, perhaps typically $e>0.9$.

In later stages of inspiral, however, the circularizing influence of
gravitational radiation reduces the eccentricity substantially.  A simple
way to estimate the eccentricity uses the constant
\begin{equation}
ae^{-12/19}(1-e^2)(1+121e^2/304)^{-870/2299}
\end{equation}
found by Peters (1964) for the lowest-order quadrupole radiation (see 
Glampedakis \& Kennefick 2002 and Glampedakis, Hughes,
\& Kennefick 2002 for a discussion of how constant this expression
is when higher-order terms are included).
Disregarding factors of
order unity, this means that $r_pe^{-12/19}$ is roughly constant.  Thus,
$r_p$ remains approximately fixed as long as $e$ is close to unity,
after which point it drops significantly.  In the first stage, where
$e\approx 1\Rightarrow a\gg r_p$, $a\propto \tau_{\rm merge}^2$ so that
$1-e=r_p/a\propto \tau_{\rm merge}^{-2}$.  Note that in this stage the
emission is strongly dominated by the time near pericenter.  Thus, 
gravitational waves are emitted for a few cycles near closest approach,
then little emission occurs until the next pericenter passage.  As a
result, the
peak frequency of the gravitational wave emission is approximately the orbital
frequency at pericenter, which is nearly constant during this
phase of inspiral.  In the second stage, 
where $a\approx r_p$ and thus $r_p\propto \tau_{\rm merge}^{1/4}$,
we have $e\propto\tau_{\rm merge}^{19/48}\approx \tau_{\rm merge}^{0.4}$.
Here the frequency of gravitational waves is roughly twice the mean
orbital frequency, or $f_{\rm GW}\propto r_p^{-3/2}$, so
$e\propto f_{\rm GW}^{-19/18}\approx f_{\rm GW}^{-1}$.  

In practice, the transition between relatively rapid change in 
eccentricity and relatively slow change happens at surprisingly
high eccentricity.  For example,
suppose the binary initially has a merger time of $10^6$~yr and an
eccentricity of 0.9.  At $10^5$~yr from merger, the eccentricity would
be $e=0.55$, at $10^4$~yr from merger
$e=0.25$, and at $10^3$~yr $e=0.1$.  If the initial eccentricity is instead
0.99, then at $10^5$~yr $e=0.83$, at $10^4$~yr $e=0.46$, and at
$10^3$~yr $e=0.2$.  The net result is that almost all the sources of
this type in the LISA band are likely to have appreciable eccentricity.
In contrast, high frequency sources detectable by ground-based instruments
will have much lower eccentricities, $e\sim 0.01$ typically.  This
means that templates based on circular orbits will be adequate for
many purposes (see Martel \& Poisson 1999 for a discussion of the
amount of signal to noise lost by nonoptimal signal processing).

A second formation scenario involves Kozai resonances in hierarchical
triple systems formed by binary-binary encounters (Miller \& Hamilton 2002b).
Here the eccentricities could be much higher and the pericenter distances
much closer than in the three-body scenario.  However, the eccentricities
will certainly not be higher than what would produce a merger within
a single orbital period of the inner binary, because the eccentricity is
increased gradually and such a short merger time would terminate the
increase.  For a merger time of $\sim 1$~yr and two $10\,M_\odot$ black
holes, this implies a pericenter distance of $\approx 10^8$~cm.  Even in
this very extreme case, by the time the binary reached the $\sim 10^7$~cm
separation necessary to bring the frequency into the LIGO-II detector
band the eccentricity would be $e<0.03$.  In more
realistic circumstances the pericenter distance would be more than
$10^9$~cm and the binary would again be nearly circular when it entered
the frequency range of ground-based detectors.

The third formation scenario involves direct two-body capture by
emission of gravitational radiation.  If this happens with a black hole
of mass $M$ and one of mass $m\ll M$, then the distance of closest
approach needed to cause a binary of initial relative velocity $v$ to
become bound is (see Quinlan \& Shapiro 1987)
\begin{equation}
\begin{array}{rl}
r_p&=\left(85\sqrt{2}\pi\over{12}\right)^{2/7}GM^{2/7}m^{2/7}(M+m)^{3/7}
c^{-10/7}v_\infty^{-4/7}\\
&\approx 7\times 10^9M_{100}^{5/7}m_{10}^{2/7}v_{\infty,6}^{-4/7}
\,{\rm cm}\; ,\\
\end{array}
\end{equation}
where $m=10m_{10}M_\odot$ and $M=100M_{100}M_\odot$.
Here $v_{\infty,6}$ is the speed at infinity in units of $10^6$~cm~s$^{-1}$.
For stellar-mass or 
intermediate-mass black holes, this distance is so much larger than
the radius $R\sim 6GM/c^2\approx 10^8M_{100}$~cm
of the last stable orbit that, again, when the frequency
is high enough to detect with ground-based instruments the orbit will
have circularized to $e<0.01$.  Note that because $r_p\propto M^{5/7}$
whereas the horizon radius $R_H\propto M$, for
supermassive black holes there is a point at which in order to
become bound the smaller object would have to pass inside the horizon,
so that a direct capture occurs.  This happens at $M\gta 2\times
10^8(v_\infty/100\,{\rm km\,s}^{-1})^{-2}\,M_\odot$, and means
that above this mass one will not observe multiple orbits.  If two-body
capture happens during a three-body interaction, the criterion is that
the capture must occur before a slight perturbation from the third
object can deflect the binary from its highly eccentric orbit.  This
means that the merger time must be less than $0.1-1$~yr.  For such
short merger times there is a significant probability that the pericenter
distance will be comparable to the radius of the last stable orbit.
This is in part because in the gravitational focusing regime the
cross section for a closest approach $r_p$ scales as $r_p$ and not
$r_p^2$.  However, we expect these to form a minor fraction of the
coalescences observed with ground-based detectors.

In summary, all of the scenarios above predict that when a
binary is orbiting with high frequency its eccentricity will be small
enough for the use of circular orbit templates.  At the frequencies
detectable with LISA, however, we expect significant eccentricities to be
typical.

\section{Rate Estimates}

What is the rate at which intermediate-mass black holes are
expected to merge with less massive compact objects?
The actual rate in a given cluster is set by the frequency of
encounters, which depends on number densities and effective
cross sections.  However, if this rate is too high then it is
self-limiting.  One upper limit is set by the
supply of smaller objects; this supply cannot be exhausted much
faster than the current age of the cluster.  Another
upper limit is set by the mass of the large black hole; the
characteristic growth time of the black hole cannot be much
shorter than the current cluster age.  We now consider these
in turn.

\subsection{Encounter rates}

The encounter rate includes contributions from two-body capture by
gravitational radiation and from three-body encounters.  Consider 
first two-body interactions.
If the smaller objects
have number densities
$10^6n_6$~pc$^{-3}$, and if they are in thermal equilibrium with
$\sim 0.4\,M_\odot$ main sequence stars with a velocity dispersion of
$10^6v_{\rm ms,6}$ cm~s$^{-1}$ (so that the velocity dispersion of
the black holes is $\approx 10^6(0.4/10m_{10})^{1/2}v_{\rm ms,6}{\rm
cm\,s}^{-1}=2\times 10^5 m_{10}^{-1/2}v_{\rm ms,6}$~cm~s$^{-1}$),
then the rate of two-body capture by a given large black hole is
\begin{equation}
\nu_{\rm enc}=\langle n\sigma v_\infty
\rangle\approx 2\times 10^{-8}
n_6m_{10}^{11/7} M_{100}^{12/7}v_{\rm ms,6}^{-11/7}\,{\rm yr}^{-1}
\end{equation}
(e.g., Quinlan \& Shapiro 1987; Miller \& Hamilton 2002a).
The typical rate can therefore be $\sim 10^{-10}-10^{-6}$~yr$^{-1}$,
depending on the cluster parameters and black hole masses.

Now consider three-body interactions. If the large black hole is initially
solitary, it will acquire a companion when it interacts strongly with an
existing binary, where ``strongly" means roughly a closest approach less
than the semimajor axis $a$ of the binary.  The interactions are dominated
by gravitational focusing, so the cross section for a close approach within
distance $a$ is roughly $\pi a (2GM/v^2)$, where $v=10^6v_6$~cm~s$^{-1}$ is
the velocity at infinity.  For a moderately hard binary with $a$=1~AU (and
thus an orbital velocity of 30~km~s$^{-1}$), this cross section is about
$2\times 10^{30}M_{100}v_6^{-2}$~cm$^2$.  If 10\% of stars are in binaries,
their number density is about $3\times 10^{-51}n_6$~cm$^{-3}$, so the
interaction time is $t=1/(n\sigma v)= 1/[(3\times 10^{-51}n_6)(2\times
10^{30}M_{100}v_6^{-2})(10^6v_6)]=5\times 10^{12}v_6M_{100}^{-1}n_6^{-1}$~s.
If the relative velocity of binaries with large black holes is somewhat
smaller than the main sequence velocity dispersion (due to the higher mass
of binaries), this is typically $\sim 5\times 10^6M_{100}^{-1} n_6^{-1}$~yr.

Once the massive black hole is in a binary, the question is how long it
will take for subsequent interactions to harden it so that it merges
because of gravitational radiation.  Since there is not a significant
loss cone for objects of this mass, the black hole can now interact
with all the stars  (density $3\times 10^{-50}n_6$~cm$^{-3}$), so at
least at first the interaction rate is about 1 per
$5\times 10^5\,M_{100}^{-1}n_6^{-1}$~yr.  The fractional hardening per
interaction is small because of the large mass ratio, but the
eccentricity can increase substantially (Quinlan 1996).  Depending on
how the eccentricity evolves, merger within a few hundred interactions
is likely to be typical.  For a  $10^3\,M_\odot$ black hole this will
give rates of $\sim 10^{-7}n_6$ mergers per year, with lower
rates for less massive black holes.  Thus, the overall rate of
interactions is likely to be dominated by three-body encounters, but
the merger rate could have a significant contribution from two-body
contributions as well because such an interaction leads to a direct
merger.

\subsection{Supply of objects}

A model-independent upper limit on the rate of mergers is set by the
requirement that the characteristic timescale on which the smaller objects
are removed from the population cannot be much shorter than the age of the
cluster, otherwise the population would be reduced rapidly.  For example,
suppose the central black hole has a mass of $10^2\,M_\odot$ and we consider
its interactions with neutron stars, of which we assume there are $10^3$
currently in the cluster (and at most a few hundred in the core).  For a
$10^{10}$~yr old cluster the interaction rate is therefore limited to not
much more than one per $10^7$~yr, and could be less.  Of course, if other
processes are more important then the rate could be much less than these
values, but the rate cannot be much more. For example, if the process of
merging with one neutron star has along the way caused the ejection of 100
neutron stars (e.g., through three-body recoil), the maximum rate drops to
one per  $10^9$~yr.  The rate is therefore $R\lta R_{\rm supply}=N_s/T_0$,
where  $N_s$ is the current number of small objects of a given type in the
cluster  and $T_0$ is the current cluster age.

\subsection{Growth rate of large black hole}

Similarly, the timescale of increase in the mass of the central black hole
cannot be much shorter than the cluster age, otherwise the massive black
hole would acquire mass rapidly and grow until it exhausted its fuel. In the
example above, if neutron stars are accreted faster than one per $10^8$~yr,
the black hole mass will grow well past its current $10^2\,M_\odot$ in a
Hubble time. Thus, $R\lta R_{\rm grow}=M/(mT_0)$, where $m$ is the mass of
the objects accreted.

This provides a guide to the most common expected interactions between an
intermediate-mass black hole and other compact objects.  Because exchange
interactions tend to favor more massive objects, the companion to an
intermediate-mass black hole is likely to be among the more massive objects
present in abundance in the core of the cluster. If there are enough
stellar-mass black holes (say, tens to hundreds), then in dense clusters
these may dominate the actual merger rate even if most of the encounters are
with other types of stars.  If the number of stellar-mass black holes is
smaller, then neutron stars (with perhaps $>10^3$ in a cluster; Grindlay et
al. 2001) or massive white dwarfs (which could constitute several
percent of the number of stars, or up to $\sim 10^4$ in a cluster) may
dominate the interactions.  As we will see in \S~4, mergers between two
black holes will provide most of the signal for high-frequency ground-based
detectors, whereas the more frequent interactions between a neutron star or
white dwarf and an intermediate-mass black hole will likely provide most of
the signal for lower-frequency space-based instruments.  Note that for
frequencies in the majority of the LISA band white dwarfs act as point
masses.  Tidal disruption occurs at the Roche separation $R_R\sim R_{\rm
WD}(M/m)^{1/3}$, where $R_{\rm WD}\sim 10^9$~cm is the radius of the white
dwarf.  This implies a gravitational wave frequency
$\sqrt{GM/R_R^3}/\pi\approx0.1$~Hz  independent of the mass of the black
hole.  Below this frequency white dwarfs can make a clean contribution to
gravitational wave signals in the LISA band.

\subsection{Overall rate of encounters}

The instantaneous rate of mergers is governed by $R_{\rm enc}$, but as the
supply is decreased and the black hole mass is increased the encounter
rate can change in response.  In addition to simply decreasing the number
of smaller objects with which the large black hole interacts, three-body
effects inject energy into the population that tend to increase the scale
height of that population.  The compact objects that interact with the
black hole will settle back via interactions with normal stars, but as the
black hole mass increases the rate of energy injection per object
increases.  This is primarily because the fractional change in binding
energy per encounter scales roughly as $m/M$ (Quinlan 1996). The binding
energy released before subsequent encounters actually eject the interloper
object therefore scales roughly as $M$.  Thus, if the initial encounter
rate is high it will decrease until it matches approximately the
characteristic time of depletion and of mass increase.

For ground-based detectors, which can see just the last few seconds at
most of a merger, it is only the overall rate that matters.  In
contrast, for space-based detectors one must also know the distribution
of time to merger for a given class of system.  To estimate this,
suppose that for a given type of secondaries one can establish the
typical time ${\bar\tau}$ between mergers.  If the probability of a
given system being a time $\tau<\tau_{\rm merge}$ away from  merger is
given by a Poisson distribution, then $P(\tau<\tau_{\rm
merge})\propto 1-\exp(-\tau_{\rm merge}/{\bar\tau})$. For example,
consider a $10^3\,M_\odot$ black hole interacting with a population of
$10^3$ neutron stars of mass $1.5\,M_\odot$.  The maximum rate of
interaction is roughly one per $10^7$~yr; if this rate is realized in a
particular cluster, the probability of that cluster having a system
less than $10^6$~yr away from merger is 10\%. The implied signal to
noise ratio for LISA, however, depends on the  eccentricity of the
system, which we examine in the next section.

Note that a young dense cluster of age $\sim 10^8$~yr has much less
serious constraints placed on interaction rates by supply limits and
mass growth timescales.  However, the maximum rate is still limited by
the encounter frequency itself.  Within $\sim 10^{8-9}$ years a massive
young cluster may be able to evolve to high density in the core,
without having yet ejected many compact objects by three-body processes
(Portegies Zwart \& McMillan 2000).  In principle, therefore, some
young clusters could have interaction rates an order of magnitude larger
than those for globulars.  To be conservative, however, we will assume
that the overall rate of mergers or inspirals from young clusters is
at most comparable to that for globulars.

\section{Signal to Noise Statistics}

\subsection{Gravitational wave amplitudes}

Suppose that a binary of eccentricity $e$ and orbital period $T$ is
a distance $r$ from us, and that in a coordinate frame in which
the center of mass of the binary is at the origin we are in a
spherical polar direction $(\theta,\phi)$, where the orbit
of the binary is in the plane $\theta=\pi/2$.  Following the notation
and development of Pierro et al. (2001), the dimensionless metric
perturbations in the $\times$ and $+$ polarizations are
\begin{equation}
\begin{array}{rl}
h_\times&={\cos\theta\over{\sqrt{2}}}\left[2h_{xy}\cos 2\phi-(h_{xx}-
h_{yy})\sin 2\phi\right]\, ,\\
h_+&={1\over{\sqrt{2}}}\left\{{3+\cos 2\theta\over{4}}\left[
2h_{xy}\sin 2\phi+(h_{xx}-h_{yy})\cos 2\phi\right]-
{1-\cos 2\theta\over{4}}(h_{xx}+h_{yy})\right\}\; .\\
\end{array}
\end{equation}
In the adiabatic approximation (i.e., the assumption that the orbital
parameters do not change significantly in a single orbit), the metric
components are expressible as sums over harmonics 
(Peters \& Mathews 1963; Peters 1964; Pierro et al. 2001)
\begin{equation}
\begin{array}{rl}
h_{xy}&=\sum_{n=1}^\infty h_{xy}^{(n)}\sin\left(n{2\pi\over T}t\right)\; ,\\
h_{xx}\pm h_{yy}&=\sum_{n=1}^\infty h_{x\pm y}^{(n)}
\cos\left(n{2\pi\over T}t\right)\; ,\\
\end{array}
\end{equation}
where the harmonic components are
\begin{equation}
\begin{array}{rl}
h_{xy}^{(n)}&=h_0n(1-e^2)^{1/2}\left[J_{n-2}(ne)+J_{n+2}(ne)-
2J_n(ne)\right]\; ,\\
h_{x-y}^{(n)}&=2h_0n\left\{J_{n-2}(ne)-J_{n+2}(ne)-2e\left[J_{n-1}(ne)-
J_{n+1}(ne)\right]+(2/n)J_n(ne)\right\}\; ,\\
h_{x+y}^{(n)}&=-4h_0J_n(ne)\; .\\
\end{array}
\end{equation}
Here the $J_n$ are Bessel functions and the prefactor is
\begin{equation}
h_0={cT\over{4\pi r}}{1-\Delta^2\over{\chi^{5/3}}}\; ,
\end{equation}
where $\Delta\equiv |M-m|/(M+m)$ is the fractional mass
difference and $\chi\equiv cT/\left[2\pi G(M+m)/c^2\right]$.
For a circular orbit, the power is all at the second harmonic
($f_{\rm GW}=2f_{\rm bin}$) and if $M\gg m$ the angle-averaged
dimensionless strain is
\begin{equation}
h\approx 7\times 10^{-21}(f_{\rm GW}/10^{-4}\,{\rm Hz})^{2/3}M_{100}^{2/3}
m_{10}(10~{\rm kpc}/r)\; .
\end{equation}
The harmonic of peak amplitude is $N_{\rm max}(e)\propto (1-e^2)^{-3/2}$
(see Pierro et al. 2001).

\subsection{Initial inspiral}

The majority of the inspiral is detectable only with low-frequency
instruments such as LISA.
Signal to noise estimates for LISA are complicated by the expectation
of a significant background of astrophysical sources, especially
white dwarf binaries (Phinney 2001).  As the detector acquires
data and frequency resolution becomes better due to the longer
observational baseline, individual sources can be identified.  They can
therefore be removed from the data stream, so that the unresolved
background will effectively diminish in strength. However, until that
point the true signal to noise is less than what would be computed from
the instrumental curve alone, particularly for $f_{\rm GW}<10^{-3}$~Hz,
where the white dwarf background is expected to make its strongest
contribution. There are also numerous other sources, especially
coalescing supermassive black holes, that may produce a strong
astrophysical background over much of the LISA frequency range (e.g.,
Phinney 2001; Hughes 2002).  As with the white dwarf sources, in the
course of the lifetime of the experiment such individual sources can be
accounted for in analysis, so that the effective sensitivity of LISA to
other sources will improve with time.  For the estimates in this
section, we consider only the expected instrumental noise curve (kindly
provided by R. Stebbins).  We also consider only the far-field inspiral.
For inspiral effects at smaller distances, including the effects of
black hole spin, see, e.g., Finn \& Thorne (2000) and Hughes (2001).

As is evident from \S~4.1,  the detectability of a signal depends
strongly on its frequency distribution and amplitude, because of the
strong frequency dependence of the detector sensitivity (e.g., Danzmann
2000).  In particular, the harmonic content of the signal is important;
orbits with high eccentricities have important contributions at
frequencies many times the orbital frequency.  For example, an orbital
frequency of $10^{-5}$~Hz is far below the most sensitive range of LISA,
but for $e=0.99$ most of the power is in the $10^{-3}-10^{-2}$~Hz range
where the S/N=5 threshold in a one year observation is a dimensionless
strain of $h_5\approx 10^{-23}$.

If the time to merger is more than about $10^6$~yr, then the signal to noise
ratio from a source at a fixed distance in a year's integration with LISA
depends on the merger time but is insensitive to other parameters such as the
individual masses or the eccentricity.  To show this, note that
$h^2f^2\propto L$, where $L\propto (GMm/a)/\tau_{\rm merge}$ is the
luminosity and  $f$ is the frequency, and that at the low-frequency end the
strain that gives a signal to noise ratio of 5 in a one year integration is
$h_5\propto f^{-5/2}$ for frequencies less than $\sim 10^{-3}$~Hz$^{-1}$ (R.
Stebbins, personal communication).  If $M\gg m$ and $1-e\ll 1$, then the
merger time is $\tau_{\rm merge}\propto a^4M^{-2}m^{-1}(1-e^2)^{7/2}$ and
the frequency of peak emission is $f_{\rm peak}\propto
(1-e^2)^{-3/2}M^{1/2}a^{-3/2}$. For simplicity, assume that all the power
in the signal is concentrated at $f_{\rm peak}$.  Then
\begin{equation}
{\rm S/N}\propto h/h_5\propto M^{-1/8}m^{-3/16}(1-e^2)^{5/32}
\tau_{\rm merge}^{-19/16}\; .\\
\end{equation}
The weak dependence on the masses and eccentricity arises because of
competing effects.  For example, at a fixed merger time a binary with
a high eccentricity will have a large semimajor axis (and thus low
fundamental frequency), but the frequency peak is at a high harmonic.
A binary with a low eccentricity will have a small semimajor axis and
thus high fundamental frequency, but the frequency peak is at a low
harmonic.  In both cases the frequency peak is about the same.
At merger times $\tau_{\rm merge}\lta 10^6$~yr, however, the signal
to noise is larger for more circular binaries because the frequency
peak is greater than or about the optimal frequency $\sim 3\times 10^{-3}$~Hz
for LISA, and the signal to noise is greater when the signal is
concentrated into few frequency bins.

The total signal to noise is the square root of the sum of the squares of
the signal to noise ratios for each of the harmonics. To compute the
signal to noise we therefore need to (1)~pick an eccentricity and merger
time, (2)~calculate the frequency distribution and dimensionless strain
amplitudes for the harmonics, (3)~take the ratio with the expected LISA
sensitivity at each frequency, and (4)~sum these in quadrature for the
final signal to noise.  The computation of a cumulative signal to noise
plot thus requires assumptions about the distribution of eccentricities
for a given merger time, although from the calculation above this only
makes a significant difference if $\tau_{\rm merge}\lta 10^6$~yr.

As a sample calculation, let us assume that for large merger time the
probability distribution of eccentricities is the thermal distribution
$P(e)=2e$ that emerges from close interactions of three equal-mass objects
(Heggie 1975).  We expect that this will underestimate the eccentricity for
large merger times, where the eccentricity may be pushed to higher values by
many small interactions (Quinlan 1996).  At merger times much less than the
typical time between encounters, the eccentricity will decrease due to
gravitational radiation.  As mentioned earlier, Peters (1964) showed
that to lowest order the eccentricity
and semimajor axis evolve so as to keep the expression
$ae^{-12/19}(1-e^2)(1+121e^2/304)^{-870/2299}$ constant.  Combining this
with the expression for merger time, we take as an approximation that when
$\tau_{\rm merge}<\tau_{\rm enc}$, the eccentricity is diminished by
$e\approx e_0(\tau_{\rm merge}/\tau_{\rm enc})^{0.4}$.  We assume that the
merger time is given by a Poisson distribution,
$P(t<t_0)\propto 1-\exp(-t_0/{\bar\tau})$, where ${\bar\tau}$ is the typical
time between mergers as defined above. Given the limits of growth of the
black hole and consumption of the secondaries, we use
${\bar\tau}=10^7$~yr for $M=10^3\,M_\odot$ and $m=1\,M_\odot$.  This is the
minimum value of ${\bar\tau}$ allowed, so if the time is greater the
contours are moved to the left in Figure~2. In all cases, we fix the
distance to the globular at 10~kpc, a representative distance to the
globulars around our Galaxy (actual distances vary from 2.2~kpc to 122~kpc,
with roughly 55\% being less than 10~kpc distant; see Harris 1996).

\centerline{\plotone{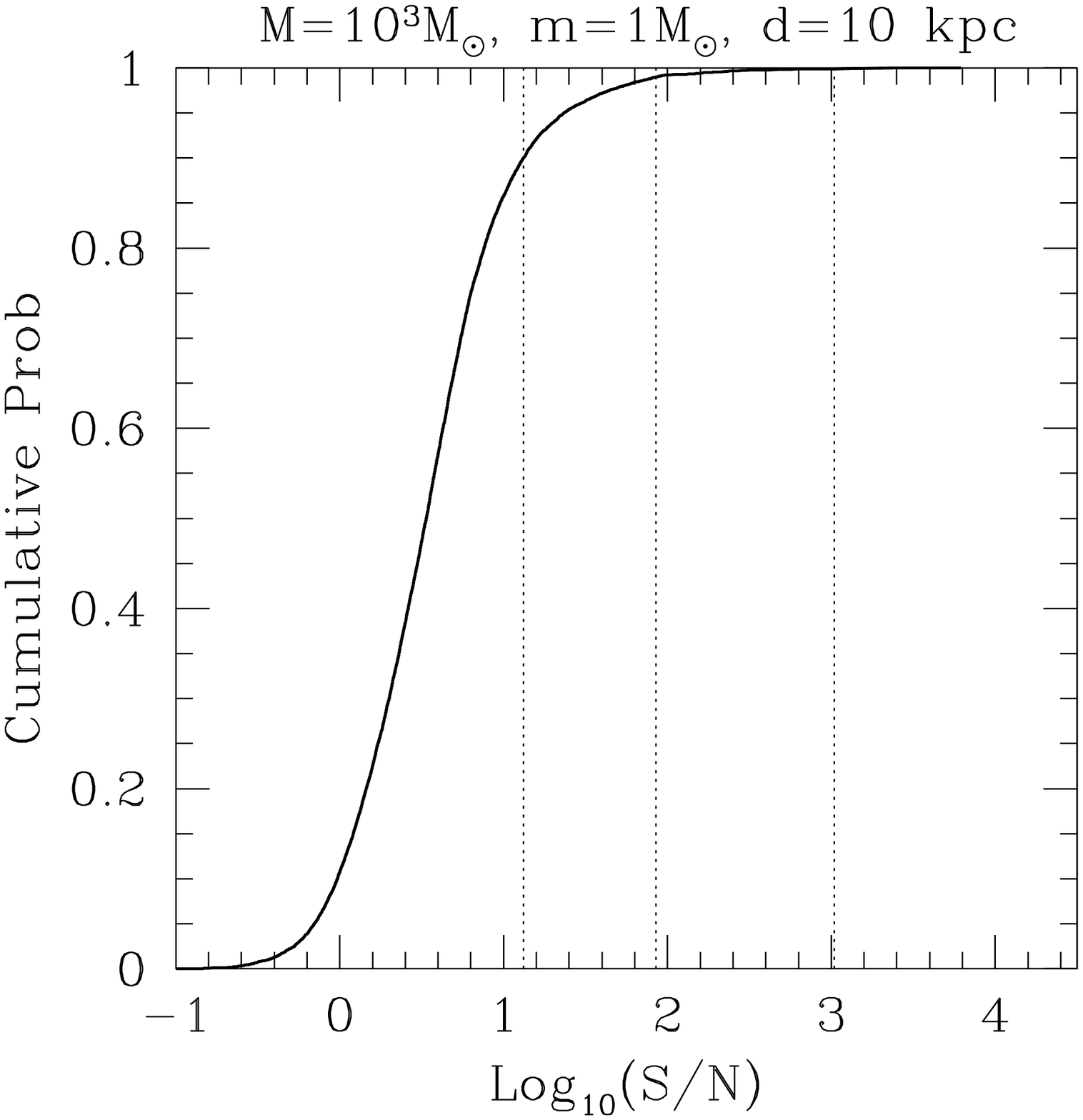}}
\vskip -0.3truein
\figcaption[]{Sample cumulative probability distribution of LISA one-year
angle-averaged signal to noise for binaries consisting of a
$10^3\,M_\odot$ black hole and a $1\,M_\odot$ compact companion (either
a neutron star or a massive white dwarf), at a distance of 10~kpc from
us.  See text for other parameters.  From this graph we see that, e.g.,
90\% of events will have S/N$>1$, 15\% will have $S/N>10$, and so on.
The vertical dotted lines are at
cumulative probabilities of 90\% (S/N$>13$), 99\% (S/N$>85$), and 99.9\%
(S/N$>1050$).  If $\sim 20$\% of
the globulars around the Milky Way have black holes of this mass and
central densities $\gta 10^5$~pc$^{-3}$, this suggests that several
sources will have one-year LISA signal to noise ratios above 10, with an
expected maximum of $\sim 20-100$.}

If these curves are roughly representative, it suggests that globulars
of at least moderate density will have a significant probability of
containing a high signal to noise gravitational wave source, especially
if $10^3\,M_\odot$ black holes are common. In a ten-year LISA
integration the signal to noise ratio would be greater than 10 for $\sim
90$\% of the Galactic globulars  that have $10^3\,M_\odot$ black holes
and a significant supply of $1\,M_\odot$ compact objects. This could be
several tens of globulars around our Galaxy.  Over ten years the
strongest of the sources could have an accumulated signal to noise of
several hundred.  There is also a chance to detect inspiraling
intermediate-mass black hole binaries at greater distances with LISA.
For example, the Virgo cluster, at a distance $\approx$16~Mpc (e.g.,
Graham et al. 1999), contains $\sim 10^3$ galaxies and $\sim 10^5$
globulars (using the mass fraction $M_{\rm gc}/M_{\rm stars}\approx
2\times 10^{-3}$ of McLaughlin 1999).  From the model above, we expect
there to be a few binaries within $\sim 10^3$~yr of merging, which would
therefore have signal to noise ratios $>5-10$ in a single year LISA
integration.  As we discuss in \S~5, such binaries are likely to have
appreciable eccentricities and detectable frame-dragging, so that there
are a number of interesting strong gravity effects that may be probed.

\subsection{Final inspiral and merger}

As discussed in \S~2, coalescing binary black holes in dense
stellar clusters may contain intermediate-mass black holes or
may consist of two stellar-mass black holes (via the Kozai mechanism
or due to a close pass in a two- or three-body interaction).  We
first treat intermediate-mass black holes, then discuss the possible
rates for stellar-mass black hole coalescence.

By the time that a binary with an intermediate-mass black hole is
in the frequency range of a ground-based instrument, it will have
shrunk through a large enough factor in pericenter distance that
the orbit will be nearly circular.  As discussed by 
Flanagan \& Hughes (1998a,b) and Cutler \& Thorne (2002),
the remaining coalescence can then be conveniently divided into
inspiral (where analytical calculations are adequate), merger
(where numerical simulations are mandatory because of strong
gravity effects), and ringdown (where analytic theory exists).
The boundary between inspiral and merger is somewhat arbitrary,
but it can be said to be roughly where the inward radial speed
increases rapidly (see Ori \& Thorne 2000).  Flanagan \& Hughes (1998a)
conservatively assume that the inspiral is ended when the orbital
frequency becomes $0.02\,c^3/(GM)$; we discuss later the effects of
spin on this number.

It is likely that radiation of significant energy and angular momentum
in a merger phase requires that the total angular momentum of the
system exceed the Kerr threshold for the total mass of the system (S.
Hughes, personal communication).  From the work of Pfeiffer, Teukolsky,
\& Cook (2000), the orbital angular momentum at the last quasicircular
orbit between two black holes (with total mass $M$ and reduced mass $\mu$) is
$\approx (2.5-3.5)\mu MG/c$, depending on the initial spin magnitudes
and orientations.  Even for equal masses, $\mu=M/4$, this is not
enough by itself to exceed the Kerr maximum of $M^2G/c$, and hence the
black holes must be prograde and spinning fairly rapidly for there to
be  a hangup phase before final merger (S. Hughes, personal
communication). For the higher mass ratio inspirals that are the main
focus here, it is probably impossible to exceed the Kerr threshold unless the
larger black hole is already close to maximally rotating.  It is
therefore likely that there will be no merger radiation per se, but
that any excess energy or angular momentum would be emitted in an
extended ringdown phase. However, in order to be complete we will
include estimates of the detectability assuming that a merger-like
phase releases an energy that we scale to a fraction
$0.1(4\mu/M)^2=\epsilon_m(4\mu/M)^2$ of the total mass-energy of the system
(following the estimate by Flanagan \& Hughes 1998a).  Similarly, the
ringdown energy is scaled to a fraction $0.03(4\mu/M)^2=
\epsilon_r(4\mu/M)^2$ of
the total mass-energy.

For a given $M$ and $\mu$, there
is a luminosity distance out to which each of these phases can be detected with
a given instrument with a signal to noise ratio of at least 10.
From Flanagan \& Hughes (1998a), for the advanced LIGO instrument
these luminosity distances for a source at redshift $z$ are
\begin{equation}
\begin{array}{rl}
D_i&\approx 3(1+z)^{-1/2}M_{100}^{-1/2}(4\mu/M)^{1/2}\,{\rm Gpc}\approx
1.8(1+z)^{-1/2}\mu_{10}^{1/2}M_{100}^{-1}\,{\rm Gpc}\\
D_m&\approx 7.6(1+z)(\epsilon_m/0.1)^{1/2}M_{100}(4\mu/M)\,{\rm Gpc}
\approx 3(1+z)(\epsilon_m/0.1)^{1/2}\mu_{10}\,{\rm Gpc}\\
D_r&\approx 0.85(1+z)^{5/2}(\epsilon_r/0.03)^{1/2}M_{100}^{5/2}
(4\mu/M)\,{\rm Gpc}
\approx 0.3(1+z)^{5/2}\mu_{10}M_{100}^{3/2}(\epsilon_r/0.03)^{1/2}\,{\rm Gpc.}\\
\end{array}
\end{equation}
The actual formulae are more complicated but these expressions are
reasonably accurate over the $50-1000\,M_\odot$ range of interest
(except for ringdown, which scales as $M$ above $240\,M_\odot$ and
$M^{-1/2}$ above $620\,M_\odot$).  The rate
of detection at S/N$>$10 is given by
\begin{equation}
R=\int {4\pi\over 3}D(M)^3\nu(M)n_{\rm GC}f(M)\,dM\; .
\end{equation}
Note that for $h=H_0/100\,{\rm km\,s}^{-1}{\rm Mpc}^{-1}\approx
0.7$, the redshift is $z=0.13$ at a distance of 2~Gpc and $z=0.4$ at
a distance of 3~Gpc, so cosmological corrections will usually be
small.
Here $n_{\rm GC}\approx 8h^3$~Mpc$^{-3}$ is the number density of globular
clusters in the local universe (as estimated by Portegies Zwart \&
McMillan 2000) and $\nu(M)$ is the rate at which smaller
objects merge with black holes of mass $M$ in a given
cluster.  The mass distribution of large black holes in clusters is
$dN/dM=f(M)$, where $\int f(M)\,dM=f_{\rm tot}<1$ is the total fraction of
globular clusters that have intermediate-mass black holes.  As a specific
example, let $\nu(M)=10^{-9}\mu_{10}^{-1}(M/100\,M_\odot)$~yr$^{-1}$ and
$f(M)=[f_{\rm tot}/\ln(M_{\rm max}/M_{\rm min})]\,M^{-1}$ for  $M$ between
$M_{\rm min}$ and $M_{\rm max}$ and zero otherwise. For $M_{\rm max}\gg
M_{\rm min}$ the approximate rates are then
\begin{equation}
\begin{array}{rl}
R_i&\approx 10h^3(f_{\rm tot}/0.1)\mu_{10}^{1/2}M_{\rm min,100}^{-2}
[\ln(M_{\rm max}/M_{\rm min})]^{-1}\,{\rm yr}^{-1}\\
R_m&\approx 60h^3(f_{\rm tot}/0.1)\mu_{10}^2M_{\rm max,100}
[\ln(M_{\rm max}/M_{\rm min})]^{-1}(\epsilon_m/0.1)^{3/2}\,{\rm yr}^{-1}\\
R_r&\approx 0.07h^3(f_{\rm tot}/0.1)\mu_{10}^2M_{\rm max,100}^{5/2}
[\ln(M_{\rm max}/M_{\rm min})]^{-1}(\epsilon_r/0.03)^{3/2}\,{\rm yr}^{-1}\; .\\
\end{array}
\end{equation}
For example, if $M_{\rm min}=50\,M_\odot$, $M_{\rm max}=300\,M_\odot$,
$h=0.7$, $\mu_{10}=1$, $\epsilon_m=0.1$, and $\epsilon_r=0.03$ then
$R_i\approx 8(f_{\rm tot}/0.1)$~yr$^{-1}$, $R_m\approx 34(f_{\rm
tot}/0.1)$~yr$^{-1}$, and $R_r\approx 0.2(f_{\rm tot}/0.1)$~yr$^{-1}$.
At a different signal to noise threshold $SNR$, these rates should be
multiplied by roughly $(10/SNR)^3$, modulo cosmological corrections.
Note that the rapid increase in number with diminishing signal to noise
ratio means that for the purposes of statistical analysis these will
dominate the constraints possible with these data.  Similar answers are
obtained for more general power laws, $f(M)\propto M^{-p}$.

Note that the division of numbers between inspiral and merger is based on
the conservative assumption that inspiral continues only to a frequency
$0.02\,M^{-1}$ (Flanagan \& Hughes 1998a), which is the frequency at the
innermost stable circular orbit for a test particle around a nonrotating
black hole. As shown in \S~2.2, the random walk process of accretion of
smaller black holes is likely to produce spin parameters of order a few
tenths. For example, a $100\,M_\odot$ black hole that has grown from
$50\,M_\odot$ by accretion of $10\,M_\odot$ black holes has a mean spin
parameter $j\approx 0.5$ (see Figure~1).  At this spin the frequency is
increased by a factor 1.7 for equatorial prograde orbits, implying
an increase of up to a factor of several in the detection rate because the
detection sensitivity increases with frequency in this range.  Further
analysis of this effect will be important.

For the LISA instrument, black holes in this mass range will not be
detectable during merger and ringdown.  From Flanagan \& Hughes (1998a),
the final inspiral could be detected at $S/N\geq 10$
in a one year integration out to a luminosity distance
\begin{equation}
D_{i,{\rm LISA}}\approx 0.2(1+z)M_{100}^{1/2}\mu_{10}^{1/2}\,{\rm Gpc}\; .
\end{equation}
Thus, the expected rate of objects detectable in a one year integration
is
\begin{equation}
R_{i,{\rm LISA}}\approx 0.02h^3(f_{\rm tot}/0.1)\mu_{10}^{1/2}
M_{\rm max,100}^{3/2}[\ln(M_{\rm max}/M_{\rm min})]^{-1}\,{\rm yr}^{-1}\;.
\end{equation}
Note, however, that over a longer integration the rate goes up dramatically
because the gravitational wave amplitude scales with frequency as
$h\sim f_{\rm GW}^{2/3}$ whereas in this frequency range the S/N=5
threshold
of LISA scales as $f_{\rm GW}$.  For example, in a 10 year integration
LISA would be expected to see several to tens of objects in the last
phase of their inspiral.  This leads to the interesting possibility that
a $50-100\,M_\odot$ black hole coalescing with a $10\,M_\odot$ black hole
may be observed with LISA a few years prior to merger, then its final
merger could be seen later with ground-based instruments.  The waveform
observed with LISA could be projected to the final merger, so that the
time, phase, and other characteristics could be anticipated and detected
with ground-based gravitational wave detectors and the region could be
observed simultaneously with conventional telescopes.  Given that the
angular resolution of LISA would at best be a few degrees for these
sources (e.g., Cutler 1998), it may be necessary to employ wide-field
monitoring to catch the final merger.  We note that detection of any
significant electromagnetic radiation during the merger would require
a profound revision of our understanding of these systems and possibly
of gravitational radiation itself.  In a globular cluster the gas density
is too low for there to be meaningful accretion from the interstellar
medium.  If at one point there was a substantial accretion disk around
either compact object, it is likely that it would have been completely
disrupted or accreted by the time of the final merger.  In addition,
gravitational radiation is not thought to couple significantly to
an electromagnetic field, so negligible photon luminosity is expected
from the final merger.  If instead there is an electromagnetic counterpart
to a merger of an intermediate-mass black hole with a neutron star or
stellar-mass black hole, parts of this picture must be revised.

Let us now turn from interactions of intermediate-mass black holes to
interactions among only stellar-mass black holes.  As before, there is
an upper limit to the rate set by the timescale in which the supply of
stellar-mass black holes is used up.  For example, if a cluster contains
$N_{\rm BH}$ stellar-mass black holes, a strong upper bound on the
binary black hole merger rate by any mechanism is $\sim 10^{-10}N_{\rm
BH}$~yr$^{-1}$ assuming that the supply is not exhausted in much less
than a Hubble time.  The actual rate depends on details.  Consider for
example merger by the Kozai mechanism after a binary-binary encounter
(Miller \& Hamilton 2002b).  The rate is proportional to the square of
the fraction of black holes in binaries. If most black holes are in
binaries then the rate could be several tens of percent of the maximum
(see Miller \& Hamilton 2002b).  If the fraction in binaries is $f_b$,
then, the rate per cluster is $\sim 10^{-10}N_{\rm BH}f_b^2$~yr$^{-1}$.
We expect that the clusters that contribute most to the rate of stellar
black hole coalescences will be those with relatively high central
densities, because low-density clusters will have a low interaction
rate.  Let us parameterize the fraction of clusters that contribute
significantly as $f$, where $f$ is likely to be a few tenths.  Note that
the denser clusters are also the ones most likely to produce
intermediate-mass black holes, hence it is likely that $f\approx f_{\rm
tot}$, but to keep the processes distinct we use different variables to
represent the fractions.  The volume  rate in the universe for mergers
of two stellar-mass black holes via the Kozai mechanism is then $\sim
10^{-2}N_{\rm BH}h^3f_b^2(f/0.1)\,{\rm Gpc}^{-3}$ yr$^{-1}$.  The other
way stellar-mass black holes can coalesce in a cluster is by having a
near approach during a three-body encounter.  As estimated in \S~6, the
rate of such coalescences will be roughly 10\% the rate at which black
holes are ejected from the cluster by dynamic recoil.  Using the
bounding estimates of Portegies Zwart \& McMillan (2001; their eqns. 8
and 9), this implies a rate in the universe of  $\sim
(1-6)h^3(f/0.1)\,{\rm Gpc}^{-3}$~yr$^{-1}$.  Thus, if $f_b>0.3$ then the
Kozai mechanism dominates, otherwise direct coalescence in three-body
interactions is more important.  From Flanagan \& Hughes (1998a), the
inspiral of a pair of $10\,M_\odot$ black holes could be detected with
S/N$\geq 10$ out to a distance 1.6~Gpc.  This implies a combined rate of
$\sim 10h^3(f/0.1)\left[10^{-2}N_{\rm BH}f_b^2+1\right]$~yr$^{-1}$ in
the advanced LIGO detector for two stellar-mass black holes, compared
with a rate of $\sim 40(f_{\rm tot}/0.1)$~yr$^{-1}$ for mergers of
stellar-mass black holes with intermediate-mass black holes.

\section{Information from Waveforms}

\subsection{Pericenter precession}

The high expected eccentricities of binaries in the LISA band
imply that it may be possible
to observe precession of the pericenter much deeper in a
gravitational well than is possible for known Galactic neutron
star binaries.  The angle of precession in an orbital period is
\begin{equation}
\Delta\phi=6\pi GM/\left[a(1-e^2)c^2\right]
\end{equation}
(e.g., Misner, Thorne, \& Wheeler 1973, pg. 1110).  If $e\approx 1$, this is
$\Delta\phi\approx 3\pi GM/(r_pc^2)$.  The effect of precession will be
to split the single frequency of the orbit into a pair 
with a separation that can be detected if the observation is for a time
$t_{\rm obs}>[r_p/(2GM/c^2)]T$, where as before $T$ is the orbital time
(cf. Pierro et al. 2001).  Combining factors, the required observation time
to barely resolve pericenter precession is
\begin{equation}
t_{\rm obs}>4m_{10}^{5/8}M_{100}^{-1/4}(1-e^2)^{-35/16}(1-e)
(\tau_{\rm merge}/10^6\,{\rm yr})^{5/8}\,{\rm yr}\; .
\end{equation}
For example, a $1\,M_\odot$ compact object in an $e=0.9$ orbit around
a $10^3\,M_\odot$ black hole, with a merger time of $10^6$~yr,
requires roughly two years of observation. When the same binary has a
merger time of $10^3$~yr (and therefore has $e\ll 1$), only two days
of observation are required.  This suggests that binaries in the Virgo
cluster, which are numerous enough for $\tau_{\rm merge} \lta 10^3$~yr
to be probable, are good candidates for observation of pericenter
precession, and that such precession may also be observable in
our own Galactic globular system.

\subsection{Lense-Thirring precession}

The Lense-Thirring precession rate, at which the axis of the orbital
plane changes, is approximately $\omega_{\rm LT}=2jG^2M^2/c^3r^3$.
Integrated over an orbit of semimajor axis $a$ and eccentricity $e$, the
average rate is $\omega_{\rm LT}=2jG^2M^2/[c^3a^3(1-e^2)^{3/2}]$. As is
evident from the formulae in \S~4.1, the signature of such precession
would be a periodic change in the relative amplitudes in the two
polarizations as the angle $\theta$ varies. For precession of more than
a radian to occur during an observation time $t_{\rm obs}$ therefore
requires
\begin{equation}
t_{\rm obs}>100 j^{-1}m_{10}^{3/4}M_{100}^{-1/2}(1-e^2)^{-9/8}
(\tau_{\rm merge}/10^6\,{\rm yr})^{3/4}\,{\rm yr}\; .
\end{equation}
Thus it will typically be difficult to detect this effect in Galactic
globulars unless the signal to noise ratio is so high that precession
of $\lta 0.1$~radians can be detected.  However, a binary in the
Virgo cluster with a $1\,M_\odot$
and a $10^3\,M_\odot$ black hole with $j=0.1$ that will merge within
$10^3$~yr will require $\approx 0.3$~yr of observation, so this is well
within reach.

An even better hope of detecting Lense-Thirring precession lies
in characterization of the inspiral waveform with a ground-based detector
such as the advanced LIGO detector.  The ratio of  Lense-Thirring to
orbital frequency is $2j(GM/rc^2)^{3/2}$ for nearly circular orbits, so
if the expected tens to hundreds of orbits of inspiral are observed then
the precession of the orbital plane should be evident.  The high
inclination orbits expected in this scenario are sensitive to the
multipole moments of the mass distribution and may therefore test the
no-hair theorem (Ryan 1996), although signal to noise ratios of tens are
typically required for significant constraints in a single inspiral (Ryan
1997).   Nonetheless, statistical combination of the constraints from
observations of multiple mergers, even at relatively low signal to noise,
could provide interesting limits.  Observations of these orbits may
therefore allow initial mapping of spacetime around rotating black holes
(particular in the final merger detectable with ground-based
instruments), a job expected to be completed with high precision by LISA
observations of stellar-mass black holes being consumed by supermassive
black holes in galactic centers (Hughes 2001).

\subsection{Decay of orbit}

For binaries close to the end of their inspiral, the orbital
frequency could change enough during $\sim 10$~yr that the
inspiral is detected.  An orbit with merger time 
$\tau_{\rm merge}$ will change its frequency by a fractional
amount $\delta\approx 0.4(t_{\rm obs}/\tau_{\rm merge})$ in an
observation of duration $t_{\rm obs}$.
For this to be detected via a one cycle shift
during the observation requires that $t_{\rm obs}\gta T/\delta$, or
\begin{equation}
t_{\rm obs}>30m_{10}^{3/16}M_{100}^{1/8}(1-e^2)^{-21/32}
(\tau_{\rm merge}/10^6\,{\rm yr})^{11/16}\,{\rm yr}\; .
\end{equation}
With some luck (specifically $\tau_{\rm merge}\lta 10^5$~yr) this
may be observable within the Galactic globular system, but once
again the Virgo cluster binaries are excellent candidates, where
at signal to noise ratios greater than 10 the decay of the orbit
will be clear in the signal.

One can combine the three effects discussed in this section to speculate
about astrophysical information that might be available from
gravitational radiation.  An interesting possibility is that the
distance to the Virgo cluster could be estimated from gravitational wave
signals alone, with an accuracy that is competitive with optical
measurements.  Alternatively, using optically measured distances as an
input, the system is overdetermined and detection of pericenter
precession and orbital decay would allow strong consistency checks of
the underlying formulae.  Suppose that LISA is operational for ten
years.  Then, from the estimates in \S~4, there will be several sources
in Virgo that are detectable with LISA with a signal to noise of
S/N$>50$ in that ten year period.  With such a high signal to noise, the
modulation due to the orbit of the Earth will localize the sources to
within several degrees, and the membership in Virgo will be based on
this association. From the discussion of black hole spin and binary
eccentricity in \S~2, the typical eccentricity for such binaries would
be $e\approx 0.1$, and the typical spin parameter would be $j\approx
0.1$.  Thus pericenter precession, Lense-Thirring precession, and decay
of the orbit would all be detectable.  The eccentricity would be evident
from the waveform. This means that the combinations $(M+m)/a^3$ (from
the frequency), $(M+m)/a$  (from pericenter precession), and
$a^4/[(M+m)Mm]$ (from decay of the orbit) are all independently
measurable.  Combining these, $a$, $m$, and $M$ would all be
determined.  Along with the strength of the waves at the detector, this
would yield an estimate of the distance to the cluster from the
gravitational radiation alone (for discussion of globular cluster
distance determinations using gravitational waves, see Benacquista
2000).   It will also be possible to estimate the spin angular momentum
of the larger black hole from Lense-Thirring precession.

\section{Conclusions}

Dense stellar clusters are promising locations for sources of
gravitational radiation.  We estimate that up to several tens of times per
year, the advanced LIGO detector will see the coalescence of a small
black hole with a larger one (a few to tens during inspiral, and
most of the rest during merger, with both numbers dependent on black
hole spin).   Mergers of two stellar-mass black holes in clusters will
likely be detectable at a rate of a few per year.
As always in astrophysical scenarios there are sources
of uncertainty. The most major is the question of the number
of black holes that are to be expected in a dense cluster, whether
in the $\sim 10\,M_\odot$ stellar-mass range or the $\sim
10^{2-4}\,M_\odot$ intermediate-mass range.  Perhaps the most easily
calculable input to this question is the number of black holes that
are originally produced.  If the initial mass function for the
cluster was the Salpeter function $dN/dM\propto M^{-2.35}$ 
above $1\,M_\odot$ and flatter below $1\,M_\odot$ (e.g., Meyer et al.
2000), then approximately 0.1\% of stars began with $M>25\,M_\odot$
and presumably evolved into black holes.  If the initial mass function
was instead the Scalo (1986) distribution, which drops off more sharply
at higher masses, the fraction with $M>25\,M_\odot$ is more like 0.05\%
(Portegies Zwart \& McMillan 2000).  Current dense clusters have $\sim
10^6$ stars, and may have had several times more at birth, so the
number of stars that evolved into black holes is typically $\sim 10^3$.

There are, however, numerous ways in which black holes may be lost from
the cluster.  The first is in the supernova that produced the black
hole. Neutron stars are known to have significant birth kicks of tens to
hundreds of km~s$^{-1}$ (e.g., Hansen \& Phinney 1997; Fryer \& Kalogera
1997).  The mechanism for this is still debated (Spruit \& Phinney 1998;
Kusenko \& Segre 1999;  Lai, Chernoff, \& Cordes 2001), but it is
thought that similar kick velocities for black holes are much smaller
(e.g., Brandt, Podsiadlowski, \& Sigurdsson 1995; Podsiadlowski et al.
2002; see Nelemans, Tauris, \& van den Heuvel 1999 for a somewhat
different perspective), if for no other reason than that black holes are
several times more massive than neutron stars so that a fixed energy or
momentum in the kick would lead to reduced speeds.  It is therefore
plausible that black holes do not receive birth kicks of $\gta
50$~km~s$^{-1}$, in which case they are retained in the cluster.

A second loss mechanism involves three-body recoil.  Several simulations
have shown that black holes of a fixed mass $\sim 10\,M_\odot$ in binaries
with other such black holes tend to be ejected by three-body interactions
before they can merge by gravitational radiation (Kulkarni et al. 1993;
Sigurdsson \& Hernquist 1993; Portegies Zwart \& McMillan 2000).  The actual
fraction of ejected black holes can depend on the mass function of stars and
other variables, but recent estimates suggest that 10\% or more of the
initial black holes can be retained by their clusters over a Hubble time
(Portegies Zwart \& McMillan 2000).  If more massive black holes are present
initially, then Miller \& Hamilton (2002a) showed that they are usually not
ejected.  These can grow by mergers after multiple three-body encounters,
but typically they will eject several to tens of field stars  along the
way.  The majority of the encounters, however, will not be with black holes so
this mechanism is not expected to deplete the black hole supply
significantly.  There are, in addition, at least two ways in which multibody
interactions can produce a merger without accompanying dynamical recoil.

One, discussed by Miller \& Hamilton (2002b), is that binary-binary
interactions can produce a stable hierarchical triple system, and if the
inclination of the orbit of the tertiary to the orbit of the inner binary is
in the right range then a secular Kozai resonance can increase the
eccentricity of the inner binary to the extent that it merges before the
next encounter, without significant recoil. The impact of this effect
depends on the binary fraction and the distribution of inclinations (see
Miller \& Hamilton 2002b), but this could allow the merger of some tens of
percent of the original population of black holes.

The other recoilless possibility involves resonant encounters in
three-body interactions, in which the three objects orbit hundreds or
thousands of times before resolving into a binary and an unbound single
star.  If, during these orbits, two black holes pass close enough to
each other that losses to gravitational radiation in a single pass cause
rapid merger, then again there is no dynamical kick. Various estimates
suggest that in an equal-mass three-body encounter the probability of
the closest approach being  less than $\epsilon<1$ times the initial
semimajor axis is $\sim \epsilon^{1/2}$ (Hut 1984; McMillan 1986;
Sigurdsson \& Phinney 1993).  Two $10\,M_\odot$ black holes must
approach to within $\sim 10^9$~cm to merge in a year, which will happen
in a given resonant encounter with a probability of a few tenths of a
percent for a semimajor axis of a few astronomical units.  If it takes
$\sim 10$ equal-mass encounters to harden a binary to the point of
ejection, this suggests that several percent of binaries will merge
before ejection in this fashion. Combining all of the above effects, it
seems likely that tens of percent of the original black hole population
will not be ejected by three-body recoil, leaving a present-day
population of hundreds.

The third loss mechanism involves the merger itself.  The emission of
gravitational waves during inspiral and merger will be somewhat asymmetric,
leading to recoil.  Calculations thus far have focused on
the weak-field limit.  They suggest that in the post-Newtonian limit the
kick scales as $a_{\rm LSO}^{-4}$ with the separation  $a_{\rm LSO}$ at
the last orbit before dynamical instability. At the separation $a_{\rm
LSO}=6GM/c^2$ appropriate for test particles around slowly rotating
black holes, the kick will be a few
kilometers per second for the mass ratio $M/m=2.6$ that maximizes the
kick (Fitchett 1983; Wiseman 1992). Binaries with large mass ratios or
nearly equal masses experience less recoil (for example, by  symmetry,
equal-mass binaries have no kick).  Also, $a_{\rm LSO}$ is greater in
a comparable-mass binary than in the test particle limit (Pfeiffer et al.
2000), which also decreases the kick.  It is not clear what level of recoil
is to be expected in the merger phase, where the radial velocity becomes
rapid.  Strong-field calculations are required to resolve whether this
process is dominant (perhaps kicking most merging black holes out of the
cluster) or insignificant (if the recoil speeds are much less than the
$\sim 50$~km~s$^{-1}$ escape speeds from the core).

Thus, if mergers do not kick black holes out of dense clusters, one may
expect at least tens to hundreds of black holes in many current systems.
These are expected to reside primarily in the core of the cluster, where
they have a greater tendency to interact with the more massive (and hence
compact) objects in the cluster.  This may explain why no definitive
examples of black hole low-mass X-ray binaries are known in the globular
cluster systems of the Milky Way or Andromeda; such a population would not
undergo mass transfer and would thus be observable only by its
gravitational wave emission. Note, however, that there is a population of
$>10^{39}$~erg~s$^{-1}$ sources in the globulars around a number of
ellipticals (Angelini, Loewenstein, \& Mushotzky 2001; White, Sarazin, \&
Kulkarni 2002; Kundu, Maccarone, \& Zepf 2002).  Possible differences
between these systems are an important subject for future study.

What about the fraction of clusters with intermediate-mass black holes?
Miller \& Hamilton (2002a) estimate that clusters with central densities
greater than $\sim 10^5$~pc$^{-3}$ have high enough encounter rates to
produce $10^{2-4}\,M_\odot$ black holes.  In the Milky Way globular system,
this would imply that roughly 40\% of globulars could host such objects
(Pryor \& Meylan 1993). However, to be conservative, we have adopted 10\%
as our fiducial value for the estimates of merger rates.  We have also
been conservative in assuming that the number density and mass of globulars
is the same out to 2-3~Gpc as it is in the local universe.  As first discussed
by Aguilar, Hut, Ostriker (1988), evaporation and tidal interactions
attenuate the globular system of a galaxy.  Therefore, it is possible that
coalescence rates a few billion years ago were higher by up to a factor
of a few than they are now (Portegies Zwart \& McMillan 2000), but this
is highly uncertain.

The general model described here is one that can be tested and enhanced
in ways both observational and theoretical.  From the observational
standpoint it is important to continue kinematic work on globulars to
look for evidence of the velocity and density cusps that are expected
signatures of black holes (Bahcall \& Wolf 1975; Frank \& Rees 1976;
see Gebhardt et al. 2000, D'Amico et al. 2002 for recent results).
Further characterization of the ultraluminous X-ray sources is also
important. For example, if a mass estimate can be obtained via radial
velocity measurements of a companion, this will shed new light on the
nature of these objects.  From the theoretical standpoint there are
several important calculations.  These include: (1)~strong-gravity
computations of the recoil speed of a black hole binary after merger,
for different mass ratios and spins, (2)~comprehensive numerical
simulations of three-body interactions with high mass ratios, to
represent intermediate-mass black holes, and (3)~detailed numerical
analysis of binary-binary encounters and the role  of the Kozai
resonance, among others.  Whatever the results of this work, there will
be significant new understanding gained on many fronts,  including the
information that can be obtained from analysis of the waveforms of the
gravitational radiation produced by black hole mergers in dense
clusters.

\acknowledgements

We thank Scott Hughes, Sterl Phinney, Steinn Sigurdsson, Kip Thorne, and
Linqing Wen for discussions about gravitational radiation and cluster
dynamics, Tuck Stebbins for discussions and for providing LISA response
curves, and Chris Reynolds and Scott Hughes for careful readings of an
earlier version of this manuscript. This paper was supported in part by 
NASA grant 5-9756 and NSF grant AST~0098436.

\end{document}